\documentclass[conference]{IEEEtran}
\IEEEoverridecommandlockouts

\usepackage{caption}

\usepackage{eucal}
\usepackage{placeins}
\usepackage{makecell}
\usepackage{svg}
\usepackage{array}
\usepackage{tabularx}
\usepackage{booktabs}
\usepackage{multirow}
\usepackage{cite}
\usepackage{amsmath,amssymb,amsfonts}
\usepackage{algorithm}
\usepackage{algpseudocode}
\usepackage{graphicx}
\usepackage{textcomp}
\usepackage{xcolor}
\usepackage{graphicx}
\usepackage{subcaption}
\def\BibTeX{{\rm B\kern-.05em{\sc i\kern-.025em b}\kern-.08em
    T\kern-.1667em\lower.7ex\hbox{E}\kern-.125emX}}

\usepackage{fancyhdr}
\usepackage{hyperref}

\fancypagestyle{firstpage}{
    \fancyhf{}

    \fancyfoot[C]{%
        \footnotesize
        \href{khaled.jebari@imt-atlantique.net}{khaled.jebari@imt-atlantique.net}
        \;\textbar\;
        \href{luiz.anetneto@imt-atlantique.fr}{luiz.anetneto@imt-atlantique.fr}
        \;\textbar\;
        \href{pyndiah.ramesh@wanadoo.fr}{pyndiah.ramesh@wanadoo.fr}
        \;\textbar\;
        \href{jl.debougrenet@imt-atlantique.fr}{jl.debougrenet@imt-atlantique.fr}
    }
}

\begin{document}


\title{High Rate Quantum Block Turbo Codes}


\author{
    \IEEEauthorblockN{
        Khaled Jebari\IEEEauthorrefmark{1},
        Luiz Anet Neto\IEEEauthorrefmark{1}\textsuperscript{,}\IEEEauthorrefmark{2},
        Ramesh Pyndiah\IEEEauthorrefmark{3},
        Jean-Louis de Bougrenet de la Tocnaye\IEEEauthorrefmark{1}
    }
    \IEEEauthorblockA{
        \IEEEauthorrefmark{1}\textit{Optics Department}, \textit{IMT Atlantique}, 655 Av. du Technopôle, 29280 Plouzané, France. \\
        \IEEEauthorrefmark{2}\textit{Lab-STICC CNRS UMR 6285}, Technopôle Brest-Iroise - CS 83818, 29238 Brest Cedex 3, France. \\
        \IEEEauthorrefmark{3}\textit{Previously with the Optics Department}, \textit{IMT Atlantique}, 655 Av. du Technopôle, 29280 Plouzané, France.
    }

}

\maketitle
\thispagestyle{firstpage}

\begin{abstract}
Quantum Block Turbo Code is an iterative decoding technique for concatenated quantum error correction codes in product form. In this paper, we study new high rate quantum block turbo codes with different constituent codes. We examine the impact of the change of codes parameters and the order of constituent codes on turbo performance.
\end{abstract}

\begin{IEEEkeywords}
Quantum Error Correction, product codes, block turbo codes, iterative decoding, quantum block turbo codes
\end{IEEEkeywords}

\section{Introduction}

Turbo codes have played a major role in classical error correction, through both convolutional \cite{397441} and block turbo codes \cite{705396}, also known as turbo product codes. Their iterative decoding schemes, which leverages both intrinsic and extrinsic information, paved the way for other performing decoding schemes allowing performances only a few tens of dB from the Shannon limit in today's systems. Quantum Serial Turbo Codes (QSTC) \cite{poulin2009quantumserialturbocodes} and Quantum Block Turbo Codes (QBTC) \cite{jebari2026quantumblockturbocodes} were introduced as the quantum counterparts of the previous classical turbo codes. 

In \cite{jebari2026quantumblockturbocodes}, we introduced QBTCs and showed their performances over a depolarizing channel using a 2D combination of simple constituent $\left[\!\left[5,1,3\right]\!\right]$ and $\left[\!\left[8,3,3\right]\!\right]$ codes. We predicted then that even better performances could be obtained with other constituent codes, with higher number of logical qubits and distance. The purpose of this paper is to study more deeply and to understand the QBTC decoding theory and its turbo potential. This work is a natural continuation of \cite{jebari2026quantumblockturbocodes}, to which we refer the reader for further background and context. For completeness, we also provide a brief overview of the general framework in Section.\ref{sec:review} to facilitate the understanding of the results presented below.

The paper is organized as follows. We start with a full analysis of the complexity of the QBTC decoding algorithm in Section.\ref{sec:compl}. We then propose some constructions of high rate QBTCs in Section.\ref{sec:high}. In Section.\ref{sec:turbo}, we study the turbo performance by varying code parameters and reversing constituent codes order. Finally, we conclude our work in Section.\ref{sec:concl}.

\section{QBTCs in a Nutshell} \label{sec:review}


Inspired by the success of iterative decoding in classical coding theory, QBTCs were recently introduced in \cite{jebari2026quantumblockturbocodes} as a quantum counterpart of classical block turbo codes. Their construction relies on combining two constituent stabilizer codes through successive row and column encoding operations, resulting in a product-like structure. The associated iterative decoder exploits the exchange of soft information between the constituent decoders to progressively refine the estimation of logical errors.

\begin{figure}[t]
    \centering
    \includegraphics[width=\linewidth]{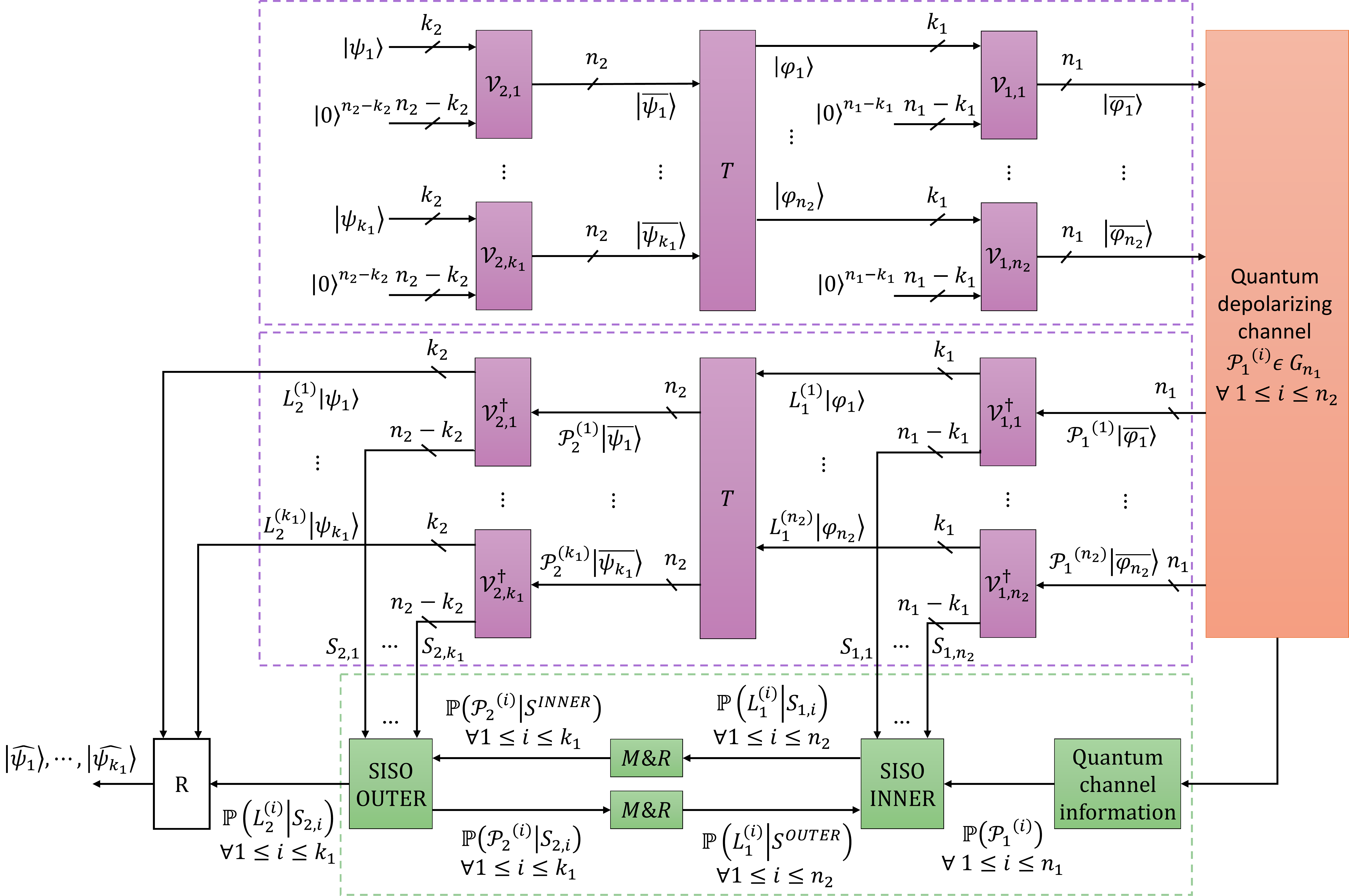}
    \caption{QBTC encoding and decoding: product code $(n_1, k_1) \times (n_2, k_2)$. Reproduced from \cite{jebari2026quantumblockturbocodes}.}
    \label{fig:nutshell}
\end{figure}

More specifically, as presented in Fig.\ref{fig:nutshell}, the decoding process alternates between two soft-input soft-output (SISO) decoders, which use syndrome information and a priori error probabilities to update the probability distributions of physical and logical errors. These distributions are exchanged through marginalization and reorganization procedures, allowing information obtained from one constituent decoder to improve the decisions of the other. By accounting for quantum degeneracy, this approach provides a flexible framework for iterative quantum error correction.

\section{Complexity Analysis} \label{sec:compl}

For a better understanding of QBTC algorithm, we consider at first a SISO unit for a code with parameters $(n,k)$. After taking the syndromes, we start by collecting the candidates only once and stored them. For all $4^k$ logical candidates $L$ and for all $2^{n-k}$ ancilla errors candidates $S$, we encode the concatenation of $L$ and $S$ using the encoding matrix $U$.

\begin{equation}
(L:S)U=\mathcal{P}.
\end{equation}

Afterwards, we fill in the logical and physical candidates sets $\Omega$ as follows:

\begin{equation}
\mathcal{P} \in \Omega(L), L \in \Omega(\mathcal{P}).
\end{equation}

The total number of pairs that needs to be considered is $4^k2^{n-k}=2^{n+k}$ which is much smaller than all $4^n=2^{2n}$ \(n\)-qubit Pauli operators. Since $U$ represents an invertible Clifford transformation, for a fixed measured syndrome:

\begin{equation}
|\Omega(\mathcal{P})|=1, |\Omega(L)|=2^{n-k} .
\end{equation}

\noindent Then, we execute the update calculations as follows:

\begin{equation}
\mathbb{P}_{upd}(L)=\mathbb{P}(L)\sum_{\mathcal{P} \in \Omega(L)}\mathbb{P}(\mathcal{P})
\end{equation}

\begin{equation}
\mathbb{P}_{upd}(\mathcal{P})=\mathbb{P}(\mathcal{P})\sum_{L \in \Omega(\mathcal{P})}\mathbb{P}(L)
\end{equation}

Consequently, once the candidate graph has been constructed, the SISO unit update algorithm have a complexity of $\Theta(2^{n+k})$. More precisely, for the logical update, it takes $4^k \times |\Omega(L)|= 2^{2k} \times 2^{n-k}=2^{n+k}$ operations. Similarly, for the physical update, it takes $2^{n+k} \times |\Omega(\mathcal{P})| = 2^{n+k}$ operations.

Now, consider two different constituent codes
$C_1 = (n_1,k_1)$ and $
C_2 = (n_2,k_2)$. There are \(n_2\) inner SISO units, each using \(C_1\), and \(k_1\) outer SISO units, each using \(C_2\). This follows directly from the row/column construction represented in \cite{jebari2026quantumblockturbocodes}. The one-time candidate-collection complexity, when candidate transformations are considered primitive, is consequently:

\begin{equation}
\Theta\left(n_22^{n_1 + k_1} + k_12^{n_2 + k_2}\right)
\end{equation}

We get the same complexity for the full SISO procedure per turbo iteration. We focus now on the M \& R  procedure, a marginalization and reorganization process that involves exchanging soft information between the inner and outer decoders by converting column probabilities into row probabilities and vice versa. Going from the inner to the outer block, the M \& R step units requires respectively a complexity of $O(k_14^{k_1})$ and $O(n_22^{n_2+k_2})$. We use $2^{n_2+k_2}$ instead of $4^{n_2}$ because the outer decoder only needs its syndrome-compatible candidate set. The same calculations can be performed going from the outer to the inner block. Thus the full cost of the M \& R procedure is: 
\begin{equation}
O\left(k_1n_2[4^{k_1} + 2^{n_2 + k_2}]\right)
\end{equation}

Therefore the complete exact decoding complexity is

\begin{equation} \label{eq:compl}
\begin{aligned}
&O\left(n_2 2^{n_1 + k_1}c_U(n_1)
+ k_1 2^{n_2 + k_2}c_U(n_2)\right) \\
&\quad + O\left(I\left(
n_2 2^{n_1 + k_1}
+ k_1 2^{n_2 + k_2}
+ k_1 n_2\left[4^{k_1} + 2^{n_2 + k_2}\right]
\right)\right)
\end{aligned}
\end{equation}

\noindent where $c_U(n)$ denotes the cost of one encoding transformation and $I$ the number of iterations. In addition, sufficient memory should be provisioned to save the different candidates


The dominant exponential factors are $2^{n_1 + k_1}$ and $2^{n_2 + k_2}$. Thus, the QBTC decoding is exponential in the constituent-code sizes, but not directly in $N=n_1n_2$. That is why using small constituent blocks remains feasible even when the final product code is considerably larger.

To circumvent this limitation, we can shrink the size of the candidate sets by eliminating those with a negligible probability. For the inner SISO, we can add restrictions on the weight of physical candidates using the code spectrum. We denote:

\begin{equation}
\Omega_{w_{\max}}(L)
=
\left\{ \mathcal{P} \in \Omega(L) : wt(\mathcal{P}) \leq w_{\max} \right\}
\end{equation}

Then, the logical update becomes:

\begin{equation}
\mathbb{P}_{upd}(L)
\approx
\mathbb{P}(L)\sum_{\mathcal{P} \in \Omega_{w_{\max}}(L)} \mathbb{P}(\mathcal{P})\approx \mathbb{P}(L) \Lambda(L)
\end{equation}

Therefore, we reduce the number of the physical candidates. In addition, for every logical candidate \(L\), we can calculate once $\Lambda(L)$ before the iterative decoding starts, which leads to $\Theta(4^{k_1})$ update complexity instead of $\Theta(2^{n_1+k_1})$.

For the outer SISO, we can do the same by cutting off physical candidates with probability under a certain threshold based on the probabilities communicated in the first iteration. We use this truncated set for the rest of the iterations. Thus, we get $\Theta(4^{k_2}m_2)$ update complexity with $m_2$ the average number of physical candidates retained per logical candidate. The truncation is efficient if $m_2<<2^{n_2-k_2}$. Unlike the inner case, this generally cannot be further reduced to $4^{k_2}$, since the physical outer probabilities change at each iteration.

We would like to mention that we could integrate the M \& R procedure into the SISO unit algorithms. In fact, the blocks could exchange marginalized probabilities instead of the full sequence probabilities. The latter could then be calculated by multiplying the input marginalized probabilities following the connection row/column as shown in Fig.\ref{fig:nutshell}. In addition, this gives us the freedom to generalize the theory to other connections between the SISO blocks.

\begin{table}[t]
\caption{Seed transformations of some QECCs.}
\vspace{-0.2cm}
\label{tab:seed codes}
\centering
\resizebox{\columnwidth}{!}{%
\begin{tabular}{>{\raggedright\arraybackslash}p{0.15\columnwidth}>{\raggedright\arraybackslash}p{0.75\columnwidth}}
\hline
\textbf{QECC} & \textbf{Seed transformation} \\
\hline
$\left[\!\left[15,7,3\right]\!\right]$ & \{876642304, 840171520, 822673408, 815923200, 623902720,
    612696064, 542310400, 8355840, 126320640, 429490176,
    715816960, 255, 3855, 13107, 21845, 26753, 25640,
    25106, 24900, 19040, 18698, 16550, 128, 2048, 8192,
    16384, 4194304, 67108864, 268435456, 536870912\}$_{10}$ \\
$\left[\!\left[11,5,3\right]\!\right]$ & \{540808,
    557073,
    589860,
    528450,
    3671040,
    603516,
    2281766,
    129840,
    1603647,
    129231,
    491583,
    49232,
    67680,
    41104,
    69792,
    1216,
    2097152,
    1024,
    1048576,
    512,
    262144,
    128\}$_{10}$ \\
$\left[\!\left[15,3,5\right]\!\right]$ & \{536891729,
    268447632,
    134238448,
    8175658,
    543785209,
    16687346,
    277316093,
    12554398,
    149979519,
    10422440,
    80412990,
    5211220,
    40206495,
    15139533,
    25264590,
    19252,
    8866,
    5737,
    536870912,
    16384,
    268435456,
    8192,
    134217728,
    4096,
    67108864,
    2048,
    33554432,
    1024,
    16777216,
    512\}$_{10}$ \\

\hline
\end{tabular}
}
\vspace{-0.5cm}
\end{table}

\section{High Rate QBTC} \label{sec:high}
\begin{figure}[t]
    \centering
    \includegraphics[width=\linewidth]{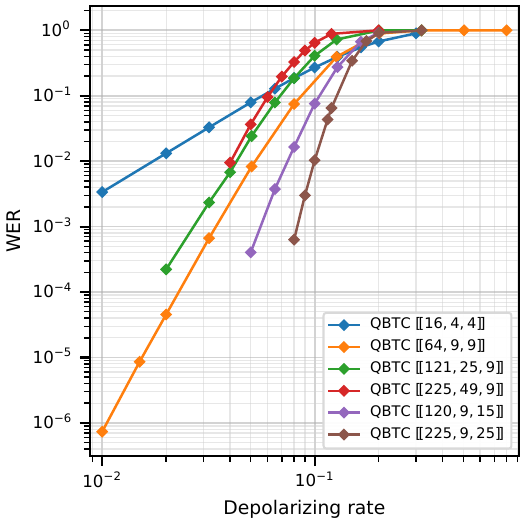}
    \caption{Comparaison of QBTC decoding performance of different code combinations from Table.\ref{tab:comb} after five iterations.}
    \label{fig:every}
\end{figure}

One of the major advantages of quantum block turbo codes is the ability to choose any constituent codes to build bigger codes with flexible code parameters. We could carefully choose those constituent codes to have a final result with relatively high code rate. We have a large catalog to choose from as presented in Table.\ref{tab:seed codes} and in \cite{jebari2026quantumblockturbocodes}. For each code, we provide the corresponding seed transformation of its encoding matrix, expressed in decimal notation. In Table.\ref{tab:comb}, we present interesting combinations of QBTCs with variant characteristics. With some careful combinations, we could reach a code rate around $0.2$ with variant code parameters like QBTC $\left[\!\left[16,4,4\right]\!\right]$, QBTC $\left[\!\left[44,10,6\right]\!\right]$, QBTC $\left[\!\left[121,25,9\right]\!\right]$, QBTC $\left[\!\left[165,35,9\right]\!\right]$ and QBTC $\left[\!\left[225,49,9\right]\!\right]$. We investigate the turbo advantage of some codes presented here in the next section. We illustrate in Fig.\ref{fig:every} the word error rate (WER)  versus depolarizing rate performance after five iterations of some code combinations from Table.\ref{tab:comb}. Their detailed assessment, per turbo decoding iteration, is provided in the next section.  

\begin{table}[b]
\centering
\caption{Combinations of QBTCs.}
\resizebox{\columnwidth}{!}{%
\begin{tabular}{c c c c c}
\hline
code & code rate & $k$ & $n$ & distance   \\
\hline
QBTC $\left[\!\left[4,2,2\right]\!\right] \times \left[\!\left[4,2,2\right]\!\right]$
 & $0,25$ & $4$ & $16$ & $4$  \\
QBTC $\left[\!\left[4,2,2\right]\!\right] \times \left[\!\left[8,3,3\right]\!\right]$
 & $0,1875$ & $6$ & $32$ & $6$ \\
QBTC $\left[\!\left[4,2,2\right]\!\right] \times \left[\!\left[11,5,3\right]\!\right]$
 & $0,227$ & $10$ & $44$ &  $6$ \\
QBTC $\left[\!\left[4,2,2\right]\!\right] \times \left[\!\left[15,7,3\right]\!\right]$
 & $0,23$ & $14$ & $60$ & $6$\\
 QBTC $\left[\!\left[8,3,3\right]\!\right] \times \left[\!\left[8,3,3\right]\!\right]$
 & $0,14$ & $9$ & $64$ & $9$ \\
 QBTC $\left[\!\left[11,5,3\right]\!\right] \times \left[\!\left[11,5,3\right]\!\right]$
 & $0,206$ & $25$ & $121$ & $9$  \\
  QBTC $\left[\!\left[15,7,3\right]\!\right] \times \left[\!\left[15,7,3\right]\!\right]$
 & $0,217$ & $49$ & $225$ & $9$  \\
  QBTC $\left[\!\left[8,3,3\right]\!\right] \times \left[\!\left[11,5,3\right]\!\right]$
 & $0,17$ & $15$ & $88$ & $9$  \\
  QBTC $\left[\!\left[8,3,3\right]\!\right] \times \left[\!\left[15,7,3\right]\!\right]$
 & $0,175$ & $21$ & $120$ & $9$  \\
  QBTC $\left[\!\left[11,5,3\right]\!\right] \times \left[\!\left[15,7,3\right]\!\right]$
 & $0,212$ & $35$ & $165$ & $9$  \\
 QBTC $\left[\!\left[4,2,2\right]\!\right] \times \left[\!\left[15,3,5\right]\!\right]$
 & $0,1$ & $6$ & $60$ & $10$  \\
 QBTC $\left[\!\left[8,3,3\right]\!\right] \times \left[\!\left[12,4,4\right]\!\right]$
 & $0,125$ & $12$ & $96$ & $12$  \\
 QBTC $\left[\!\left[11,5,3\right]\!\right] \times \left[\!\left[12,4,4\right]\!\right]$
 & $0,151$ & $20$ & $132$ & $12$  \\
 QBTC $\left[\!\left[15,7,3\right]\!\right] \times \left[\!\left[12,4,4\right]\!\right]$
 & $0,155$ & $28$ & $180$ & $12$  \\
 QBTC $\left[\!\left[8,3,3\right]\!\right] \times \left[\!\left[15,3,5\right]\!\right]$
 & $0,075$ & $9$ & $120$ & $15$  \\
 QBTC $\left[\!\left[11,5,3\right]\!\right] \times \left[\!\left[15,3,5\right]\!\right]$
 & $0,09$ & $15$ & $165$ & $15$  \\
 QBTC $\left[\!\left[15,7,3\right]\!\right] \times \left[\!\left[15,3,5\right]\!\right]$
 & $0,093$ & $21$ & $225$ & $15$  \\
  QBTC $\left[\!\left[15,3,5\right]\!\right] \times \left[\!\left[15,3,5\right]\!\right]$
 & $0,04$ & $9$ & $225$ & $25$  \\
 
\hline
\end{tabular}%
}

\vspace{2mm}

\label{tab:comb}
\end{table}

\section{Turbo Analysis}\label{sec:turbo}

The results presented in \cite{jebari2026quantumblockturbocodes} have proven that QBTCs with more than one logical qubit demonstrate reasonable turbo effect, i.e. a WER vs depolarizing rate with pronounced slope. In this section, we study such turbo performance under the change of code parameters as well as the order of the constituent codes in the encoding and decoding procedure. It is worth mentioning that, unlike classical product codes, the final code does not necessarily have valid codewords along each row and column, as some stabilizers may anticommute. Consequently, decoding must be performed in the reverse order of encoding: if encoding proceeds from rows to columns, decoding must proceed from columns to rows, and vice versa.

\subsection{Code Parameters Variation}

\begin{figure}[t]
    \centering
    \includegraphics[width=\linewidth]{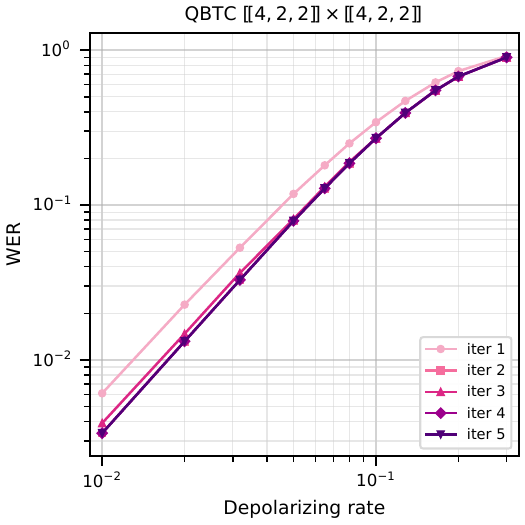}
    \caption{Performance of QBTC $\left[\!\left[16,4,4\right]\!\right]$. }
    \label{fig:1644}
\end{figure}

The first application of the QBTC decoding theory was on codes with only one logical qubit. This produced no discernible signs of turbo effect \cite{jebari2026quantumblockturbocodes}. The increase of the number of logical qubits, however, led to the emergence of the turbo effect. This observation provides motivation for studying the effects of code parameters variation on the turbo performance.

We start with the smallest code that we have with more than only one logical qubit, which is $\left[\!\left[4,2,2\right]\!\right]$. We note that due to its distance, this constituent code alone is not capable of correcting errors but only detecting them. As shown in Fig.\ref{fig:1644}, we observe that the turbo gain saturates relatively fast after the second decoding iteration. It is worth noting that for this particular code, we observe that the third iteration might provide worse results than the second, which we believe is related to its small distance $2$. This odd behavior doesn't happen to other codes with distances greater than $3$. 

\begin{figure*}[!t]
    \centering

    \begin{subfigure}{0.49\textwidth}
        \centering
        \includegraphics[width=\linewidth]{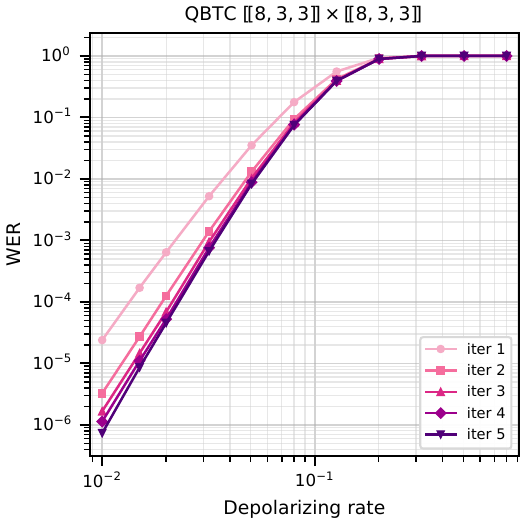}
        \caption{        \hspace{1cm} INNER  $\left[\!\left[8,3,3\right]\!\right]$ OUTER $ \left[\!\left[8,3,3\right]\!\right]$}
    \end{subfigure}
    \hfill
    \begin{subfigure}{0.49\textwidth}
        \centering
        \includegraphics[width=\linewidth]{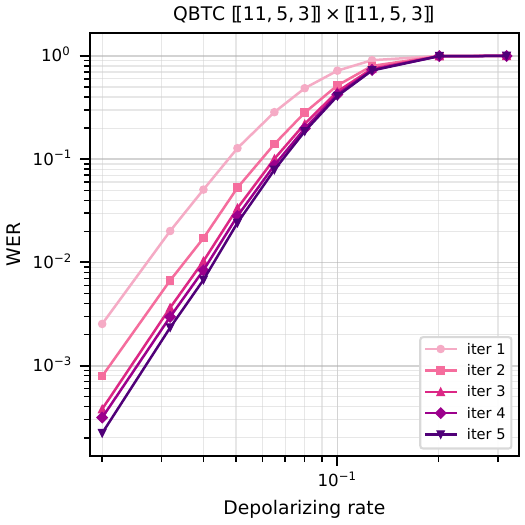}
        \caption{        \hspace{1cm} INNER  $\left[\!\left[11,5,3\right]\!\right]$ OUTER $ \left[\!\left[11,5,3\right]\!\right]$}
    \end{subfigure}

    \vspace{0.5cm}

    \begin{subfigure}{0.49\textwidth}
        \centering
        \includegraphics[width=\linewidth]{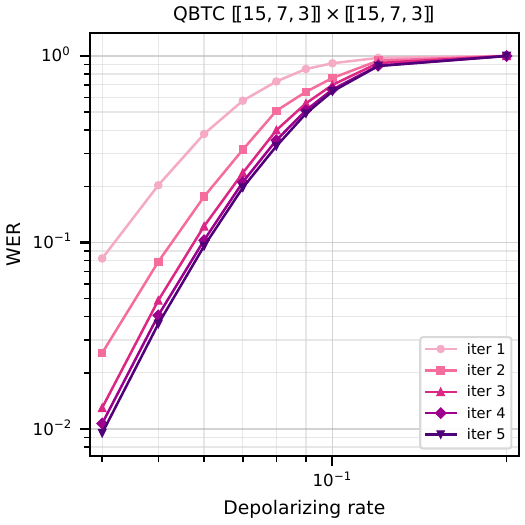}
        \caption{        \hspace{1cm} INNER  $\left[\!\left[15,7,3\right]\!\right] $ OUTER $ \left[\!\left[15,7,3\right]\!\right] $}
    \end{subfigure}
    \hfill
    \begin{subfigure}{0.49\textwidth}
        \centering
        \includegraphics[width=\linewidth]{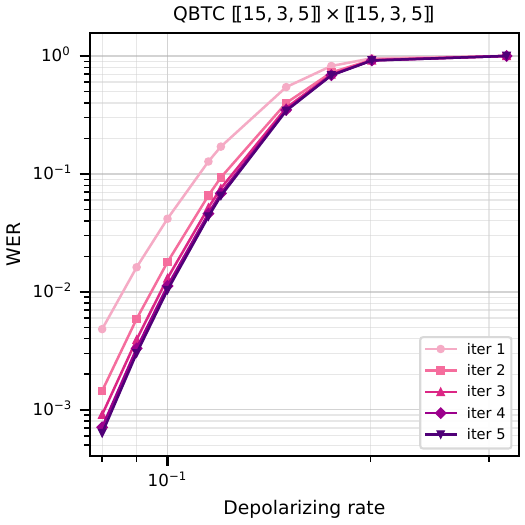}
        \caption{        \hspace{1cm} INNER  $\left[\!\left[15,3,5\right]\!\right]$ OUTER $ \left[\!\left[15,3,5\right]\!\right]$}
    \end{subfigure}

    \caption{Performance of QBTCs with variable number of logical qubits.}
    \label{fig:log qubits}
\end{figure*}

 Fig.\ref{fig:log qubits} presents results for various constituent codes, with code distances ranging from 3 to 5 and the number of logical qubits varying from 3 to 7. An overall view of these results clearly shows the increasing turbo effect when increasing number of logical qubits, as predicted. We notice that for all these codes, a large advantage is obtained between the first and the second iterations. More interestingly, we observe that, when we increase the number of logical qubits, we notice a clear improvement between the second and the third iteration. This observation is clearly illustrated in the performance of QBTC $\left[\!\left[11,5,3\right]\!\right] \times \left[\!\left[11,5,3\right]\!\right]$ and QBTC $\left[\!\left[15,7,3\right]\!\right] \times \left[\!\left[15,7,3\right]\!\right]$.
In addition, the increase of the distance of the constituent codes does appear to have visible impact on the turbo performance. The performance of QBTCs $\left[\!\left[64,9,9\right]\!\right]$ and $\left[\!\left[225,9,25\right]\!\right]$, which are two codes with the same number of logical qubits and different distance, shows a similar overall shape with a lower word error rate (WER) for the second code thanks to the higher distance. Further experiments can be done on QBTCs with an interesting code rate using $\left[\!\left[12,4,4\right]\!\right]$ and QBTCs with higher distance using $\left[\!\left[24,3,7\right]\!\right]$ or $\left[\!\left[25,5,7\right]\!\right]$ as constituent codes \cite{Grassl:codetables}.

In the previous experiments, we use QBTCs with two identical constituent codes. A more relevant property to investigate is to work with QBTCs with different constituent codes for the inner and the outer parts. The following section will elaborate on this point with the purpose of determining whether the order of constituent codes has an effect on the turbo performance.

\subsection{Constituent Codes Order}
We begin with an interesting combination that brings together two constituent codes, one of which has only one logical qubit. This case warrants further investigation, as previous studies have identified this specific code as the cause of the absence of turbo effect. 

\begin{figure*}[!t]
    \centering

    \begin{subfigure}{0.49\textwidth}
        \centering
        \includegraphics[width=\linewidth]{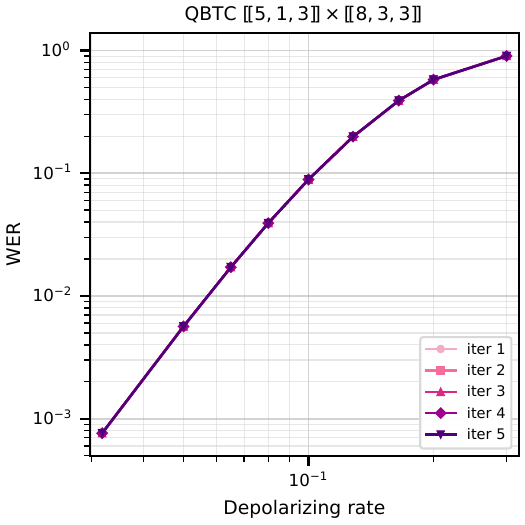}
        \caption{        \hspace{1cm} INNER  $\left[\!\left[5,1,3\right]\!\right]$ OUTER $ \left[\!\left[8,3,3\right]\!\right]$}
    \end{subfigure}
    \hfill
    \begin{subfigure}{0.49\textwidth}
        \centering
        \includegraphics[width=\linewidth]{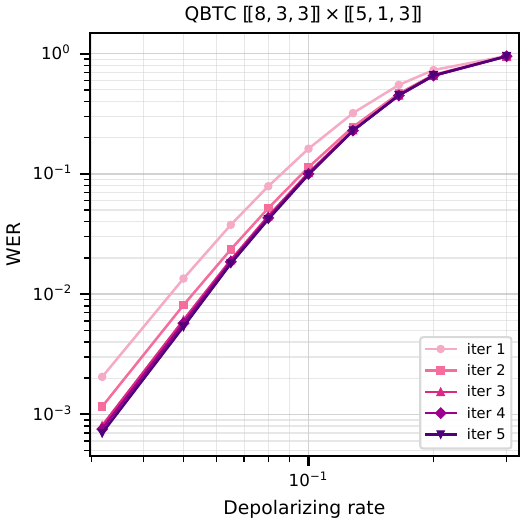}
        \caption{        \hspace{1cm} INNER  $\left[\!\left[8,3,3\right]\!\right]$ OUTER $ \left[\!\left[5,1,3\right]\!\right]$}
    \end{subfigure}

    \caption{Comparaison of the performance of QBTC $\left[\!\left[5,1,3\right]\!\right] \times \left[\!\left[8,3,3\right]\!\right]$ and QBTC $\left[\!\left[8,3,3\right]\!\right] \times \left[\!\left[5,1,3\right]\!\right]$ .}
    \label{fig:513833}
\end{figure*}

In Fig.\ref{fig:513833}, we study the case of pairing $\left[\!\left[5,1,3\right]\!\right]$ and $\left[\!\left[8,3,3\right]\!\right]$ in different orders. We clearly observe the lack of turbo effect in the case where we use $\left[\!\left[5,1,3\right]\!\right]$ in the inner block, closest to the channel. The turbo effect reappears in reverse order converging at the end to the same performance. This behavior can be explained by the fact that, at the initialization of the turbo decoding process, information from the channel first passes through the inner decoder before being propagated to the remaining components of the decoding structure. The iterative process then begins, meaning that the initial decisions made by the inner decoder can significantly influence the subsequent exchange of information and the overall decoding convergence. In this sense, the initial decoding stage plays a crucial role in shaping the iterative process. Consequently, when $\left[\!\left[5,1,3\right]\!\right]$ was employed as the inner code, no significant turbo gain was observed, as its initial decisions dominated the decoding process, limiting the ability of subsequent iterations to refine or correct them. 


Therefore, the choice of the inner code is crucial for the turbo performance. Another thing to note is that the complexity of the decoding changes with the order of the constituent codes. Following the Eq.\ref{eq:compl}, an order of magnitude estimate of the cost of an iteration of the first ordering is $8\times2^6+1\times2^{11}+1\times8\times[4^1+2^{11}]=18976$ and the second one $5\times2^{11}+3\times2^6+3\times5\times[4^3+2^6]=12352$. Thus, while the first order gives us a one shot decoding, with faster convergence with respect to the iterations, it is more complex numerically. These estimates, as we see later, scale with the size of constituent codes and will observe a remarkable difference in the complexity when changing the order of the constituent codes. The order must therefore be chosen based on the desired performance, the qubits' coherence time, and hardware availability. 

\begin{figure*}[!t]
    \centering

    \begin{subfigure}{0.49\textwidth}
        \centering
        \includegraphics[width=\linewidth]{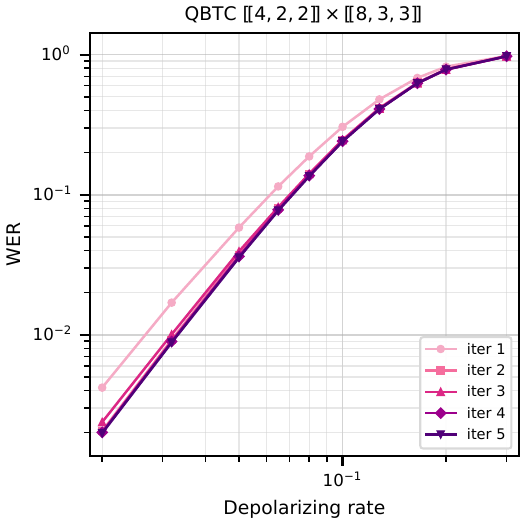}
        \caption{        \hspace{1cm} INNER  $\left[\!\left[4,2,2\right]\!\right]$ OUTER $ \left[\!\left[8,3,3\right]\!\right]$}
    \end{subfigure}
    \hfill
    \begin{subfigure}{0.49\textwidth}
        \centering
        \includegraphics[width=\linewidth]{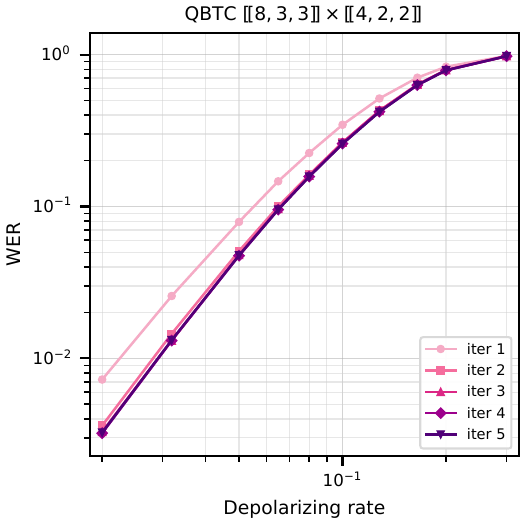}
        \caption{        \hspace{1cm} INNER  $\left[\!\left[8,3,3\right]\!\right]$ OUTER $ \left[\!\left[4,2,2\right]\!\right]$}
    \end{subfigure}

    \caption{Comparaison of the performance of QBTC $\left[\!\left[4,2,2\right]\!\right] \times \left[\!\left[8,3,3\right]\!\right]$ and QBTC $\left[\!\left[8,3,3\right]\!\right] \times \left[\!\left[4,2,2\right]\!\right]$.}
    \label{fig:422833}
\end{figure*}

\begin{figure*}[!t]
    \vspace{-3cm}
    \centering

    \begin{subfigure}{0.49\textwidth}
        \centering
        \includegraphics[width=\linewidth]{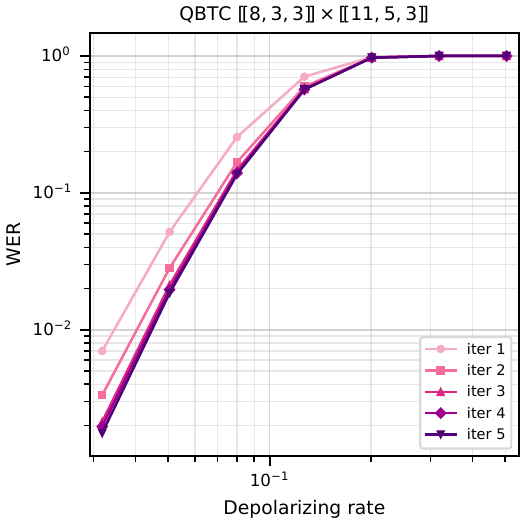}
        \caption{        \hspace{1cm} INNER  $\left[\!\left[8,3,3\right]\!\right]$ OUTER $ \left[\!\left[11,5,3\right]\!\right]$}
    \end{subfigure}
    \hfill
    \begin{subfigure}{0.49\textwidth}
        \centering
        \includegraphics[width=\linewidth]{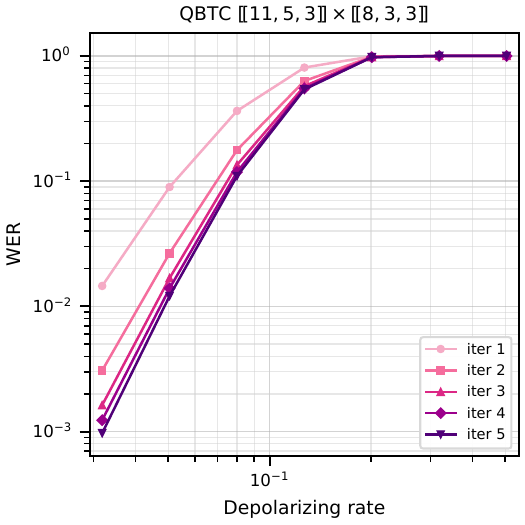}
        \caption{        \hspace{1cm} INNER  $\left[\!\left[11,5,3\right]\!\right]$ OUTER $ \left[\!\left[8,3,3\right]\!\right]$}
    \end{subfigure}

    \caption{Comparaison of the performance of QBTC $\left[\!\left[8,3,3\right]\!\right] \times \left[\!\left[11,5,3\right]\!\right]$ and QBTC $\left[\!\left[11,5,3\right]\!\right] \times \left[\!\left[8,3,3\right]\!\right]$.}
    \label{fig:8331153}
\end{figure*}

\begin{figure*}[!t]
    \vspace{-5cm}
    \centering

    \begin{subfigure}{0.49\textwidth}
        \centering
        \includegraphics[width=\linewidth]{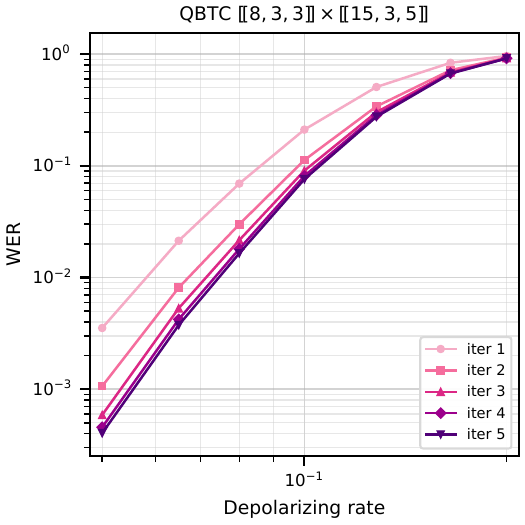}
        \caption{        \hspace{1cm} INNER  $\left[\!\left[8,3,3\right]\!\right]$ OUTER $ \left[\!\left[15,3,5\right]\!\right]$}
    \end{subfigure}
    \hfill
    \begin{subfigure}{0.49\textwidth}
        \centering
        \includegraphics[width=\linewidth]{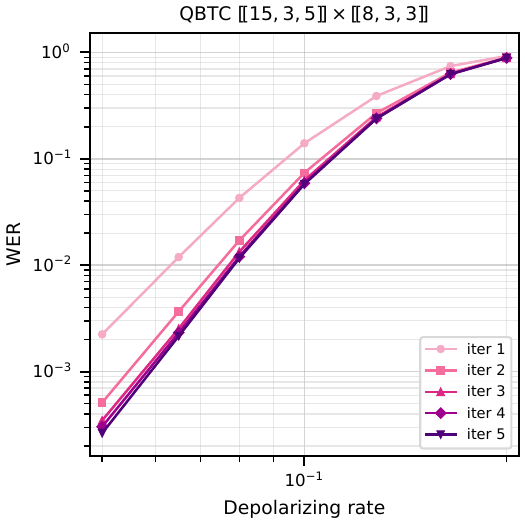}
        \caption{        \hspace{1cm} INNER  $\left[\!\left[15,3,5\right]\!\right]$ OUTER $ \left[\!\left[8,3,3\right]\!\right]$}
    \end{subfigure}

    \caption{Comparaison of the performance of QBTC $\left[\!\left[8,3,3\right]\!\right] \times \left[\!\left[15,3,5\right]\!\right]$ and QBTC $\left[\!\left[15,3,5\right]\!\right] \times \left[\!\left[8,3,3\right]\!\right]$.}
    \label{fig:8331535}
\end{figure*}

We generalize now this interpretation to other constituent codes with variable sizes through Fig.\ref{fig:422833}, Fig.\ref{fig:8331153} and Fig.\ref{fig:8331535}. An overall view over these figures shows that in general when the smaller code is used as the inner code we get a relatively faster convergence. When we inverse the order, the performance starts off slightly behind but catches up progressively with a more remarkable turbo effect. In the end, we believe that for a considerably high number of iterations the two ordering will converge to the same performance.

We detail now the complexity of the combination of the constituent codes $\left[\!\left[8,3,3\right]\!\right]$ and $\left[\!\left[11,5,3\right]\!\right]$. For the first arrangement, we get an order of $11\times2^{11}+3\times2^{16}+3\times11\times[4^3+2^{16}]=2383936$ per iteration. However, for the other arrangement, we get an order of $8\times2^{16}+5\times2^{11}+5\times8\times[4^5+2^{11}]=657408$ per iteration, which is approximately 3.5-fold lower then the first order. In a nutshell, we observe that placing the larger code in the inner block requires more Turbo Boost iterations to achieve convergence, but results in lower computational complexity per iteration. Conversely, placing the smaller code in the inner block leads to faster convergence, at the cost of higher computational complexity per iteration.

\section{Conclusion}\label{sec:concl}

This paper provides a detailed assessment of the performance of QBTC decoding starting with a full decoding complexity analysis. It demonstrated that it is exponential in the
constituent-code sizes, not directly in $N = n_1n_2$. As a result, working with relatively small constituent blocks remains practical, even when they are combined to form a much larger product code. This construction allows us to build relatively high rate QBTCs with a code rate approaching $0.2$. An analysis was conducted to evaluate the performance of these codes and examine how their performance varies with changes in the constituent codes parameters and order.
We conclude that increasing the number of logical qubits enhances the turbo advantage, resulting in greater performance gains across iterations. Furthermore, this improvement is accompanied by a reduction in the Word Error Rate (WER), particularly at larger code distances. In addition, we observe that placing the larger code in the inner block requires a stronger turbo boost, but results in lower overall complexity. Conversely, using the smaller code as an inner code leads to faster convergence, at the cost of increased complexity per iteration.

\section{LLM Usage Disclaimer}
Large Language Models (LLMs) were used in this work as a coding assistance tool to help optimize our simulation code, with the goal of improving execution speed and enabling efficient parallelization across both Central Processing Unit (CPU) and Graphics Processing Unit (GPU). All generated suggestions were reviewed, adapted, and validated by the authors. Furthermore, the obtained results were cross-checked against those produced by independently developed, manually written code to ensure their accuracy and reliability.

\bibliographystyle{IEEEtran}
\bibliography{references}

\end{document}